\documentclass[10pt,twocolumn,letterpaper]{article}

\usepackage{iccv}
\usepackage{times}
\usepackage{epsfig}
\usepackage{graphicx}
\usepackage{svg}
\usepackage{amsmath}
\usepackage{amssymb}
\usepackage{multirow}
\usepackage{graphics}

\usepackage{scalerel}
\usepackage{tikz}
\usetikzlibrary{svg.path}

\definecolor{orcidlogocol}{HTML}{A6CE39}
\tikzset{
  orcidlogo/.pic={
    \fill[orcidlogocol] svg{M256,128c0,70.7-57.3,128-128,128C57.3,256,0,198.7,0,128C0,57.3,57.3,0,128,0C198.7,0,256,57.3,256,128z};
    \fill[white] svg{M86.3,186.2H70.9V79.1h15.4v48.4V186.2z}
                 svg{M108.9,79.1h41.6c39.6,0,57,28.3,57,53.6c0,27.5-21.5,53.6-56.8,53.6h-41.8V79.1z M124.3,172.4h24.5c34.9,0,42.9-26.5,42.9-39.7c0-21.5-13.7-39.7-43.7-39.7h-23.7V172.4z}
                 svg{M88.7,56.8c0,5.5-4.5,10.1-10.1,10.1c-5.6,0-10.1-4.6-10.1-10.1c0-5.6,4.5-10.1,10.1-10.1C84.2,46.7,88.7,51.3,88.7,56.8z};
  }
}

\newcommand\orcidicon[1]{\href{https://orcid.org/#1}{\mbox{\scalerel*{
\begin{tikzpicture}[yscale=-1,transform shape]
\pic{orcidlogo};
\end{tikzpicture}
}{|}}}}

\usepackage{caption}
\usepackage{subcaption}

\usepackage[pagebackref=true,breaklinks=true,letterpaper=true,colorlinks,bookmarks=false]{hyperref}

\iccvfinalcopy 

\def\iccvPaperID{****}

\ificcvfinal\fi

\begin{document}

\title{Deep Generative Networks for Heterogeneous Augmentation of Cranial Defects}

\author{Kamil Kwarciak \orcidicon{0000-0002-1392-4291} \\
AGH University of Krakow\\
Krakow, Poland\\
{\tt\small kkwarciak@student.agh.edu.pl}
\and
Marek Wodzinski \orcidicon{0000-0002-8076-6246} \\
AGH University of Krakow, Department of Measurement and Electronics\\
Krakow, Poland\\
University of Applied Sciences Western Switzerland (HES-SO Valais), Information Systems Institute \\
Sierre, Switzerland\\
{\tt\small wodzinski@agh.edu.pl}
}

\maketitle
\ificcvfinal\thispagestyle{empty}\fi

\begin{abstract}
The design of personalized cranial implants is a challenging and tremendous task that has become a hot topic in terms of process automation with the use of deep learning techniques. The main challenge is associated with the high diversity of possible cranial defects. The lack of appropriate data sources negatively influences the data-driven nature of deep learning algorithms. Hence, one of the possible solutions to overcome this problem is to rely on synthetic data. In this work, we propose three volumetric variations of deep generative models to augment the dataset by generating synthetic skulls, i.e. Wasserstein Generative Adversarial Network with Gradient Penalty (WGAN-GP), WGAN-GP hybrid with Variational Autoencoder pretraining (VAE/WGAN-GP) and Introspective Variational Autoencoder (IntroVAE). We show that it is possible to generate dozens of thousands of defective skulls with compatible defects that achieve a trade-off between defect heterogeneity and the realistic shape of the skull. We evaluate obtained synthetic data quantitatively by defect segmentation with the use of V-Net and qualitatively by their latent space exploration. We show that the synthetically generated skulls highly improve the segmentation process compared to using only the original unaugmented data. The generated skulls may improve the automatic design of personalized cranial implants for real medical cases.
\end{abstract}

\begin{figure}[h]
\begin{center}
    \includegraphics[width=\columnwidth]{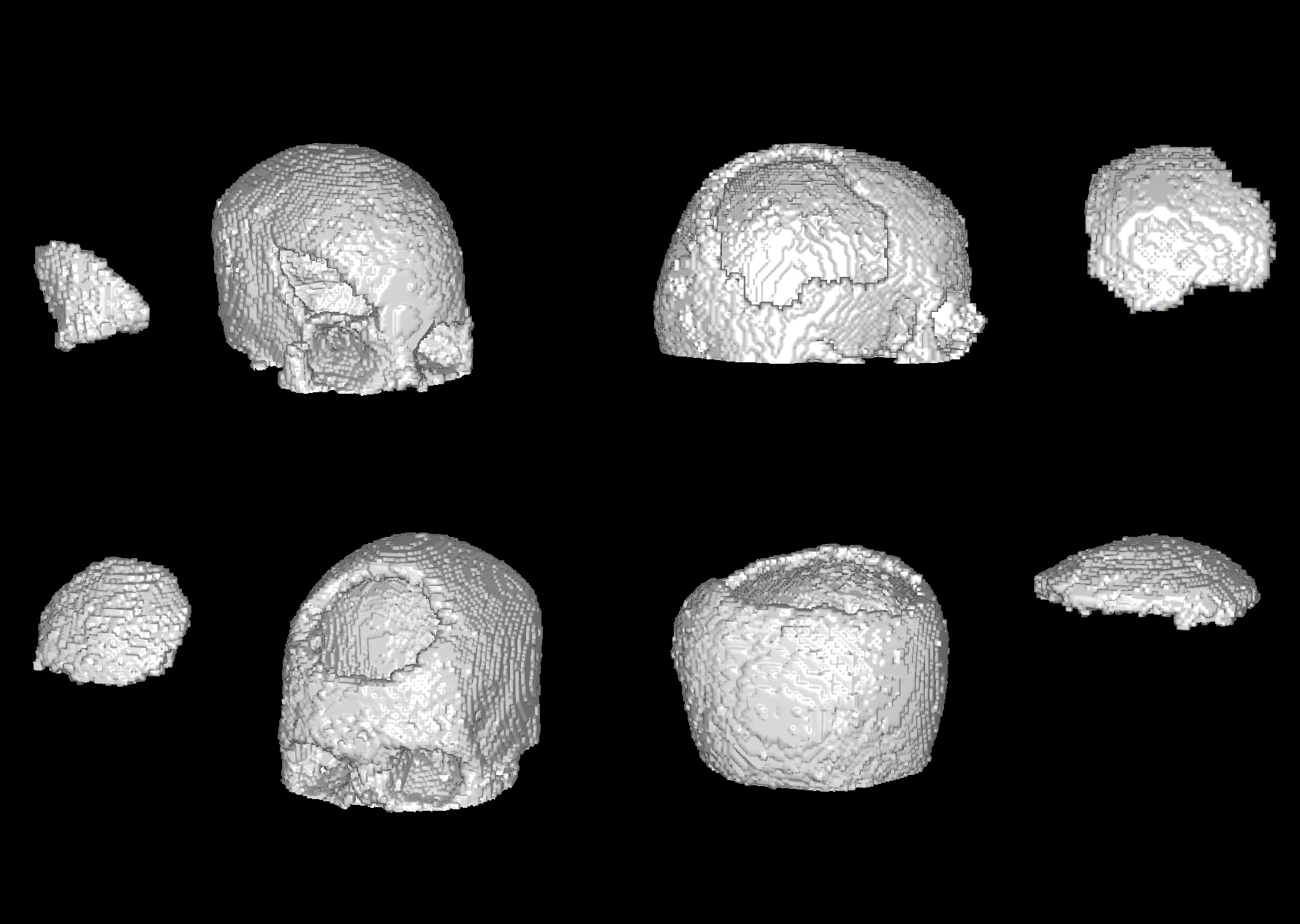}
    \caption{\label{fig:synthetic-skulls}Examples of synthetically generated defective skulls with their compatible defects with the use of VAE/WGAN-GP.}
\end{center}
\end{figure}

\section{Introduction}

Cranioplasty is a surgical action where either bone defect or deformity is repaired with the use of cranial implants. The range of differences among cranial defects seems immeasurable due to irregularities in pathological conditions, especially in size, shape, and position. The main objective related to cranioplasty from the engineering perspective is the way the implants are designed. Two main concerns of the design pipeline are data gathering and the computational design procedure itself. When it comes to design techniques, most commonly it is performed with computer-aided design (CAD) methods. However, the process is rather complicated and time-consuming. It requires professional software tools and the experience of the designer. On the other hand, in opposition to the expertise-driven CAD techniques, great success has been achieved by the data-driven deep learning approaches~\cite{ellis2020deep, Mahdi2021, LI2023102865}. However, their main difficulty is associated with the first step of the design pipeline, which is the lack of data. It originates from the fact that most of the available datasets of defective skull scans are scarce and the general process of acquisition is rather time-consuming. The amount of possible cranial defects is infinite, while the amount of patients with these diseases is highly limited. Hence, one of the directions to overcome these problems lies in the capabilities of augmenting the existing datasets by using generative models to synthesize new training samples.

In recent years, generative deep learning has strongly influenced the world of synthetic data, especially the domain of computer vision. Generative algorithms impact task-specific areas of research, such as medical imaging~\cite{Celard23}. Moreover, implant design automation is broadly impacted by the growth and scalability of deep learning algorithms. Due to huge demands for cranial implants, deep learning-based solutions arose to be a crucial part of the automation processes~\cite{ellis2020deep, Mahdi2021, LI2023102865}. However, deep learning methods require huge amounts of data to work effectively. This is one of the main challenges current approaches face as datasets of defective skulls in computer tomography (CT) or magnetic resonance (MR) modalities are relatively limited. This leads to a question: can the defective skulls and the respective implants be further augmented by deep generative models?

\textbf{Contribution: } In this work, we address the above question by developing a method that can generate dozens of thousands of unique skulls with heterogenous defects. Furthermore, we evaluate the proposed approach by performing the automatic defect reconstruction, where synthetic data significantly improves the results compared to only using the unaugmented data. We confirm that the synthetically generated data improves the model's generalizability.

\begin{figure*}[t]
\begin{center}
    \includegraphics[width=0.9\textwidth]{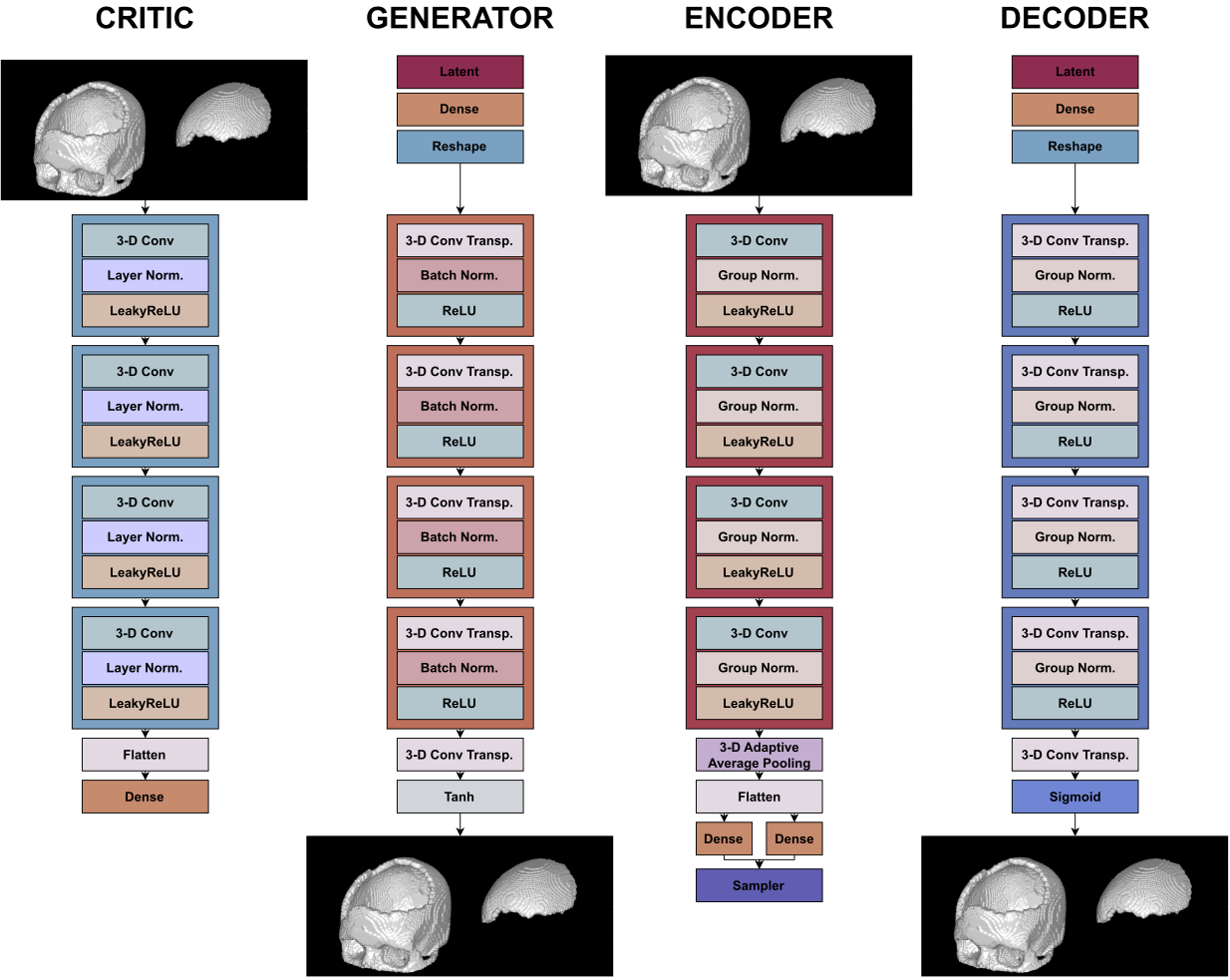}
    \caption{\label{fig:iccv-architectures}Detailed architectural building blocks for WGAN-GP, VAE/WGAN-GP, and IntroVAE. IntroVAE uses an encoder as the inference model part and a decoder as the generator part, however without sigmoidal nonlinearity.}
\end{center}
\end{figure*}

\section{Related Work}

\paragraph{Automatic cranial implants design}
Novel approaches for automatic cranial implant design strongly rely on deep learning capabilities, as most of them use encoder-decoder architectures and their variations~\cite{Li2020, Bayat2020, Mahdi2021, Pathak2021}. Other techniques are based on the nature of the data used which is due to the fact that CT scans of skulls have binary format. This leads to a significant imbalance between informative voxels that represent the bone and the uninformative voxels that represent the background, thus techniques such as sparse convolutional neural networks (CNNs)~\cite{SparseCNN2022} are found to be valuable. Finally, data augmentation techniques turned out to be highly beneficial for the whole design process, as it commonly occurs in deep learning-based solutions. Starting from the simple operations, such as permutation, scaling, and translation, and going into more advanced ones, such as inter-patient skull registration~\cite{ellis2020deep}, to finally point out the ones that use synthetic data created with generative networks~\cite{WODZINSKI2022107173}.

\paragraph{3D Generative models} 
Generative Adversarial Networks (GANs)~\cite{goodfellow2014} and Variational Autoencoders (VAEs)~\cite{kingma2014autoencoding} have been successfully applied to many research areas such as image synthesis~\cite{Park2019}, text generation~\cite{Wang2018}, audio generation~\cite{Caillon2021}, object detection~\cite{Siva2021}, pose estimation~\cite{Yang2018} and others. Hence, with their ability to learn to generate complex patterns and features, these techniques have also shown great potential in 3D modeling and generation tasks, such as creating realistic 3D objects~\cite{3dgan, HoloGAN2019}. Potential architectures and design flows have to be chosen concerning the 3D data representations, as they may differ between tasks. The most widely used representations are point clouds, meshes, neural fields, and voxel grids. The last one is most widely used in medical data representations, however, applying generative models for volumetric medical data poses additional difficulties, as they have to face very strict requirements of highly advanced accuracy and reliability, and what's most important, extremely detailed spatial structures~\cite{Kwon2019}. To overcome both potential issues, the architecture design and training pipeline has to be carefully established. For example, in the field of fluid dynamics, Xie \textit{et al.}~\cite{Xie2018} proposes two discriminators, spatial and temporal, as in this setup the model can precisely capture details of turbulent flows. The approach presented by Guan \textit{et al.}~\cite{Guan2020} addresses the concept of latent manifolds meaningfulness and is a sort of hybrid that combines Autoencoder loss with Generalized Autoencoder loss, together with Chamfer distance to create a robust volumetric generative model. Moreover, ~\cite {Kwon2019} also builds their model in a hybrid manner by combining $\alpha$-GAN with an additional code discriminator and Wasserstein GAN to generate synthetic 3D magnetic resonance images of the brain.

\paragraph{Generative models in synthetic cranioplasty}
The idea of creating synthetic skulls with the use of deep learning techniques is still only partially explored, as most of the approaches are focused on generating implants rather than whole skulls. Furthermore, even the GAN-based or VAE-based methods are not as popular in this field as the U-Net-based or pure encoder-decoder ones~\cite{Li2020, Pathak2021}. The work presented in~\cite{Pimentel2020} combines statistical shape modeling with GANs to produce cranial implants. The GAN component of the framework is trained on 2D slices of complete skulls and its generator part produces anatomically-plausible labels when 2D slices of defective skulls are inputted. SSM is responsible for defect localization and reconstruction, and GAN has a role as a postprocessing tool for defect refinement. The work that directly targets the use of generative models to obtain synthetic skulls is the one presented in~\cite{WODZINSKI2022107173}. In this work, the VAE model serves as a crucial component of the implant generation pipeline, playing a key role in generating additional 100,000 training instances of defective skulls and their compatible implants. Nevertheless, the authors do not explore different types of generative networks.

\section{Methods}

In this work, we propose three robust generative networks able to produce highly detailed defective skulls with compatible defects that can positively influence the process of automatic cranial implant design based on the U-Net framework. Firstly we present a volumetric variation of Wasserstein GAN  with Gradient Penalty (WGAN-GP)~\cite{Arjovsky2017, Gulrajani2017}, then we compose its hybrid with VAE, by creating a training routine in which the WGAN-GP as an input takes a latent manifold generated by the VAEs encoder part and finally, we introduce a volumetric version of Introspective VAE (IntroVAE)~\cite{Huang2018}. 

\subsection{Wasserstein GAN with Gradient Penalty}
Wasserstein GAN (WGAN)~\cite{Arjovsky2017} and specifically its Gradient Penalty form (WGAN-GP)~\cite{Gulrajani2017} is a highly robust model with extensive generative capabilities which overcomes typical problems occurring in GAN training pipelines such as mode collapse or vanishing gradient. The core concept behind the WGAN is to redefine the loss function by basing it on Earth-Mover ($EM$) distance also called Wasserstein-1 ($W$) distance. By definition, it is formulated as:
\begin{equation} \label{eq:3.1}
\setlength{\jot}{10pt}
    W(p_{r}, p_{g}) = \inf_{\gamma \in \Pi(p_{r}, p_{g})} {\mathbb{E}_{(x,y)\sim{\gamma}}{[\: \parallel x - y \parallel \:]}},
\end{equation}
where $\Pi(p_{r}, p_{g})$ is a set of all joint distributions between real data distribution $p_{r}$ and distribution of generated examples $p_{g}$. By applying this formulation to GANs methodology WGAN loss is meant to measure the $W$ distance between these distributions:
\begin{equation} \label{eq:3.2}
\setlength{\jot}{10pt}
    W(p_r, p_g) = \max_{w \in W}{\mathbb{E}_{x \sim{p_r}}{[f_{w}(x)]} - \mathbb{E}_{z \sim{p(z)}}{[f_{w}(g_{\theta}(x))}]},
\end{equation}
where we assume that $f$ is from the parameterized by $w$ family of 1-Lipschitz functions ${f_{w \in W}}$. This yields to the final formulation of so-called critic $C$ that computes a distance between $p_r$ and $p_g$. The smaller this distance gets, the closer the generated examples of $p_g$ are to $p_r$. What's crucial in the training framework of WGAN is the attainment of Lipschitz continuity. Its enforcement is classically achieved through gradient clipping or penalization. Gradient penalty is a more powerful approach as it provides higher training stability. Thus, the final formulation of the loss function for WGAN-GP training has a form of:
\begin{equation} \label{eq:3.3}
\begin{gathered}
    \mathcal{L} = \mathbb{E}_{\tilde{x} \sim{p_g}}{[C(\tilde{x})]} - \mathbb{E}_{x \sim{p_r}}{[C(x)]} \\ + \lambda \; \mathbb{E}_{\hat{x} \sim{p_{\hat{x}}}}{[(\parallel \nabla_{\hat{x}}{C(\hat{x})} \parallel_{2} - 1)^2]},
\end{gathered}
\end{equation}
where $p_{\hat{x}}$ is a distribution sampled uniformly along straight lines between distributions $p_{r}$ and $p_{g}$, which can be done as a linear interpolation between data from these two distributions.

\subsection{VAE pretraining of WGAN-GP}

In the training framework of WGAN-GP and many other GANs, an input to the generator is a random noise of some distribution, most commonly Gaussian or uniform. On the other hand, in the models that follow the encoder-decoder manner like VAE, a generative process performed by the decoder works on the output of the encoder part. Thus, as the whole architecture has insight into the real data distribution and both encoding and decoding components work jointly, a latent manifold created in the model's bottleneck is a highly accurate low-dimensional representation of the data, VAEs tend to generate coherent and meaningful data more quickly than GANs. However, VAEs have their own problems, like blurry samples generation and lack of diversity in the generated data. Following that we combine the strengths of both VAE and WGAN-GP, i.e. VAE accurate and informative latent manifold and WGAN-GP robust generator, to create a hybrid framework of VAE/WGAN-GP. Hence we compose a three-step training pipeline of this hybrid model. Firstly, short-term training of the VAE is performed to obtain an outline of the encoder’s low-dimensional manifold. In the second step, short-term training of the WGAN-GP is conducted, where the mentioned VAE’s latent manifold serves as a generator’s input for just a few learning epochs. Finally, WGAN-GP is trained in its standard manner, with sampling from a latent normal distribution. However, it now has a preliminary knowledge of the desired real data distribution, and its training processes exhibit enhanced robustness, stability, and improved speed.

\subsection{Introspective VAE}

Finally, we dive further into hybrid models and compose a volumetric variation of Introspective VAE (IntroVAE)~\cite{Huang2018}. In comparison to common hybrids of GANs and VAEs that use 3 components, i.e. encoder, discriminator, and decoder (or generator)~\cite{Dumoulin2017, Makhzani2016}, IntroVAE uses only encoder (the inference model) and generator. The training pipeline of the IntroVAE framework is done as follows: the encoder part of VAE serves as the discriminator of GAN and the generator (decoder) part of VAE serves as the generator of GAN. What’s more, the adversarial properties of GANs are maintained as the encoder and generator are trained jointly. This setup allows the model to discriminate between the real and generated data and generate as realistic data as possible. IntroVAE's loss captures all requirements of both VAE and GAN as it combines ELBO objective and minimax game. Encoder loss has a form of:
\begin{equation} \label{eq:3.4}
\begin{gathered}
    \mathcal{L}_{E} = D_{KL}(q_{\phi}(z|x) \parallel p(z)) \\ + \alpha \sum_{s = rec, p} 
    {\max{(0, m - D_{KL}(q_{\phi}(ng(z_{s})|x_{s}) \parallel p(ng(z_{s}))))}} 
    \\ - \beta \; \mathbb{E}_{q_{\phi}(z|x)}{[\log{p_{\theta}(x_{rec}|z)]}},
\end{gathered}
\end{equation}
where $D_{KL}$ stands for Kullback-Leibler divergence, $rec$, and $p$ indexes indicate reconstruction and new samples respectfully, $ng(\cdot)$ term stands for no backpropagation of gradients at that point, $m$ is a positive margin and $\alpha$ and $\beta$ are weighting parameters. The generator loss is defined as:
\begin{equation} \label{eq:3.5}
\begin{gathered}
    \mathcal{L}_{G} = \alpha \sum_{s = rec, p} D_{KL}(q_{\phi}(z_{s}|x_{s}) \parallel p(z_{s}))) \\ - \beta \; \mathbb{E}_{q_{\phi}(z|x)}{[\log{p_{\theta}(x_{rec}|z)]}}.
\end{gathered}
\end{equation}
IntroVAE has the property of stable training and no mode collapse occurrences, thus it is one of the most robust hybrid architectures.

\section{Experiments}

\subsection{Dataset}

We use the SkullBreak dataset \cite{Kodym2021}, which is adapted from a public head CT collection CQ500 \cite{CQ500} and consists of 114 skulls for training and 20 skulls for testing. We motivate this selection by its high heterogeneity and diversity. A total of 570 training cases and 100 testing cases are created by introducing five defects for each skull independently. The defect types are as follows: bilateral, frontoorbital, parietotemporal, random of type 1, and random of type 2. Data has a binary format where 1 corresponds to the skull and 0 corresponds to the background. The original data resolution is 512 x 512 x 512 however, for the purposes of this work, and to match the requirements of memory usability and training efficiency, the skulls are downsampled to a lower resolution of 128 x 128 x 128. It is found that this resolution still captures most of the structural and geometrical properties, and provides a better computational performance. We note that the original 512 x 512 x 512 resolution, would admittedly improve the results, however, the purpose of the work is to examine the impact of the augmentation itself, hence the impact of using higher resolution may be a subject of further work.

\subsection{Implementation details}
To implement the WGAN-GP, we utilize volumetric CNNs to construct both the critic and the generator. The critic is built with the use of 4 convolutional blocks composed of 3D convolution layers followed by layer normalization and LeakyReLU, which decrease the resolution and increase the number of channels. Its input is a 2-channel volumetric representation of the skull, meaning it has a shape of 128 x 128 x 128 x 2 where the 1st channel represents a defective skull and the 2nd channel represents its compatible defect. The generator mirrors the critic, but in its convolutional blocks, 3D transpose convolution layers are used, followed by batch normalization and ReLU nonlinearity, and it increases the spatial resolution together with decreasing the number of channels. As an input, it takes a 200-dimensional latent vector of Gaussian distributed values. To compose the VAE/WGAN-GP relation we build the VAE component similarly to the WGAN-GP, as convolutional blocks of the encoder are the same as the ones in the critic. However, group normalization is used instead of layer normalization, also in the decoder. The output of the bottleneck further proceeds into the generator part of the WGAN-GP. Finally, the architectural design of the volumetric IntroVAE is also similar to its predecessors in terms of convolutional blocks, as the inference model mimics the VAEs encoder and the generator mimics the VAEs decoder. We show the detailed architecture of every network component in the Figure \ref{fig:iccv-architectures}.

\subsection{Experimental setup}
All the code related to this work is implemented in Python with the main use of the TensorFlow framework~\cite{Abadi2016}. The training setup for each model is presented as follows. WGAN-GP is trained with the ADAM optimizer ~\cite{Kingma2014} with $\beta_{1}$ equal to 0.5, $\beta_{2}$ equal to 0.9, learning rate set to 2 $\cdot$ 10$^{-4}$ and gradient penalty $\lambda$ set to 100. Additionally based on the settings from~\cite{Arjovsky2017, Gulrajani2017} for every iteration of the generator, there are 5 iterations of the critic. For the VAE/WGAN-GP, the VAE component is also trained with ADAM optimizer and its standard hyperparameters, i.e. $\beta_{1}$ equal to 0.9, $\beta_{2}$ equal to 0.999 and learning rate of 10$^{-3}$. The training procedure lasts for 10 epochs to obtain an outline of the latent manifold and then it is fed to the WGAN-GP. In the second stage, WGAN-GP is trained for 15 epochs by feeding a generator with VAEs reparametrized latent vector. Finally, the training procedure is continued in the classic manner of WGAN-GP as discussed previously. For the IntroVAE we also use Adam with standard hyperparameters, however, we carefully select the rest of the model's hyperparameters, i.e. $\alpha$, $\beta$ and $m$ from the Equations \ref{eq:3.4} and \ref{eq:3.5}. As suggested in~\cite{Huang2018}, firstly the model is trained for just a couple of epochs in the VAE manner ($\alpha$ = 0) to find the most convenient values for $\beta$ and $m$, where $m$ should be a little larger than Kullback-Leibler divergence value of VAE. Thus, we experimentally find the following values: $\alpha$ = 0.25, $\beta$ = 1.0, and $m$ = 10.0. Finally, we also train a classic VAE with the same setup of ADAM optimizer and training flow as the one in the preliminary stage of IntroVAE and treat it as a baseline model.

All the models were trained until convergence was determined by the quality of the generated samples. A batch size of 8 was used consistently across all setups. All experiments were conducted on the NVIDIA Tesla V100 GPU.

\subsection{Postprocessing}
Additionally, we perform small three stages of postprocessing to polish some imperfections that might have occurred in the generative process. Firstly, as the default data type of neural networks’ weights is a single-precision floating-point, the desired binary skulls’ data format is not always perfectly affordable. Thus, additional binarization is applied to ensure a zero-one data character. Secondly, to ensure separability between the defective skull and the defect the following morphological operation is used:
\begin{equation} \label{eq:4.1}
    I = XOR(I, AND(I, S)),
\end{equation}
where $I$ is a generated defect and $S$ is a generated defected skull. Finally, as some generative networks might sometime produce small artifacts in the form of voxel grains or dust, it is an appropriate approach to remove them. One of the ways of achieving that is by performing connected components analysis. With the use of it objects that have a fewer number of voxels than a predefined value are simply removed.


\section{Evaluation}
With the use of pretrained generative networks described in the previous chapters, we generate 40 000 skulls per each model (including baseline VAE). This results in a total of 120 000 pairs of defective skulls and their compatible defects.

\begin{table}[htbp]
\resizebox{\columnwidth}{!}{%
\begin{tabular}{ccccc}
\hline
Model name                   & Instances & \begin{tabular}[c]{@{}c@{}}Train \\ Synthetic data\end{tabular} & \begin{tabular}[c]{@{}c@{}}Validation \\ SkullBreak - train\end{tabular} & \begin{tabular}[c]{@{}c@{}}Test\\ SkullBreak - test\end{tabular} \\ \hline
\multirow{4}{*}{WGAN-GP}     & 500       & 0.957                                                           & 0.745                                                                    & 0.612                                                            \\
                             & 1000      & 0.970                                                           & 0.784                                                                    & 0.675                                                            \\
                             & 3000      & 0.939                                                           & 0.817                                                                    & 0.727                                                            \\
                             & 40000     & 0.941                                                             & 0.898                                                                      & 0.796                                                             \\ \hline 
\multirow{4}{*}{VAE}         & 500       & 0.934                                                           & 0.513                                                                    & 0.398                                                            \\
                             & 1000      & 0.922                                                           & 0.551                                                                    & 0.410                                                            \\
                             & 3000      & 0.873                                                           & 0.771                                                                    & 0.628                                                            \\
                             & 40000     & 0.832                                                             & 0.895                                                                      & 0.728                                                              \\ \hline 
\multirow{4}{*}{VAE/WGAN-GP} & 500       & 0.961                                                           & 0.751                                                                    & 0.660                                                            \\
                             & 1000      & 0.935                                                           & 0.797                                                                    & 0.703                                                            \\
                             & 3000      & 0.890                                                           & 0.833                                                                    & 0.757                                                            \\
                             & 40000     & 0.941                                                             & 0.867                                                                      & 0.782                                                              \\ \hline 
\multirow{4}{*}{IntroVAE}    & 500       & 0.974                                                           & 0.718                                                                    & 0.589                                                            \\
                             & 1000      & 0.916                                                           & 0.784                                                                    & 0.656                                                            \\
                             & 3000      & 0.937                                                           & 0.822                                                                    & 0.678                                                            \\
                             & 40000     & 0.934                                                             & 0.926                                                                      & 0.751                                             \\ \hline  \hline

\multirow{1}{*}{Without Augmentation}    & 570       & -                                                           & 0.928                                                                    & 0.742    \\ \hline

\end{tabular}
}
\caption{\label{tab:evalDice}Mean Sørensen–Dice coefficient obtained on different sizes of datasets along with different origins of data and visibility in training processes: Train (synthetically generated), Validation (seen by generative model and unseen by V-Net), Test (unseen by generative model and unseen by V-Net).}
\end{table}

\subsection{Quantitative analysis - defect reconstruction}
From the quantitative perspective of the analysis, synthetic skulls should be able to be used in the same tasks as the real ones. In the field of cranial implant design, a crucial part is the process of segmenting the defect from the defective skull to design a suitable implant. For this part of the evaluation, we use the volumetric segmentation network V-Net~\cite{Milletari2016} in a slightly shallower variant. For each model, a subset of 500, 1 000, 3 000, and 40 000 instances are used to train the V-Net, where the input to the network is a defective skull and the desired output is a segmented defect. The objective of V-Net's training is to minimize a Dice loss defined as:
\begin{equation} \label{eq:5.1}
    L_{Dice} = 1 - 2 \cdot \frac{A \cap B}{A + B + \epsilon},
\end{equation}
where $A$ is a segmented defect, $B$ is a ground truth defect and $\epsilon$ is a smoothing factor. This approach should provide an overview of the capabilities of the synthetic datasets as in theory the larger the number of instances, the broader the dynamics and diversity should be captured. In other words, it is expected that the larger and more heterogeneous the dataset is, the greater the similarity of the synthetic distribution to the real, underlying distribution. The pretrained V-Nets are used for defects segmentation on real data of the SkullBreak dataset~\cite{Kodym2021}. We highlight three types of data: (i) train, which are the skulls synthetically generated by the previously mentioned techniques (used to train the V-Nets); (ii) validation, which are the real, training cases of SkullBreak, hence they have been seen by generative networks in their training processes, but they have not been seen by the V-Nets themselves; (iii) test, which are the testing cases of SkullBreak and they have not been seen in the training processes of either generative models or V-Nets. Following that, the evaluation comes down to computing the Sørensen–Dice coefficient expressed as:
\begin{equation} \label{eq:5.2}
    DSC = 2 \cdot \frac{I \cap D}{I + D},
\end{equation}
between the real implants compatible with the defects $I$ and segmented defects $D$, both upsampled to the original 512 x 512 x 512 resolution of SkullBreak. The results of this evaluation are presented in the Table \ref{tab:evalDice} concerning the characteristics of the data, together with the results of the V-Net segmentation trained only on the SkullBreak train data.

\begin{figure}[]
	\centering
    \begin{subfigure}{\columnwidth}
		\centering
		\includegraphics[scale=0.5]{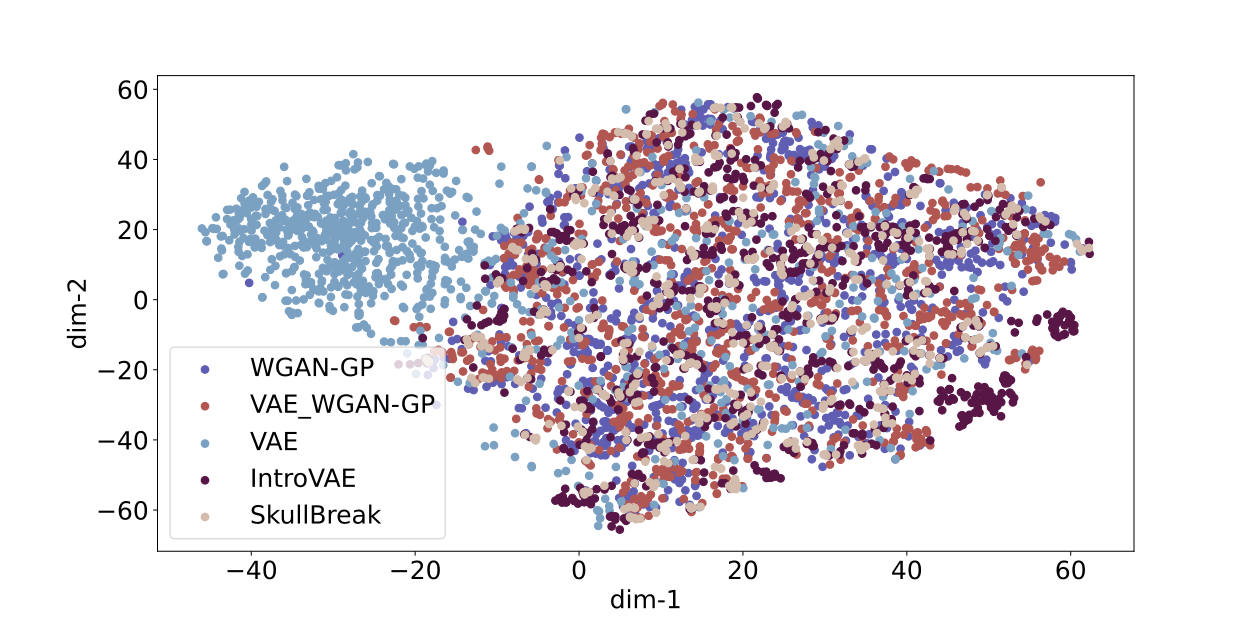}
		\subcaption{\label{tsne}}
	\end{subfigure}
	
    \begin{subfigure}{\columnwidth}
		\centering
        \includegraphics[scale=0.5]{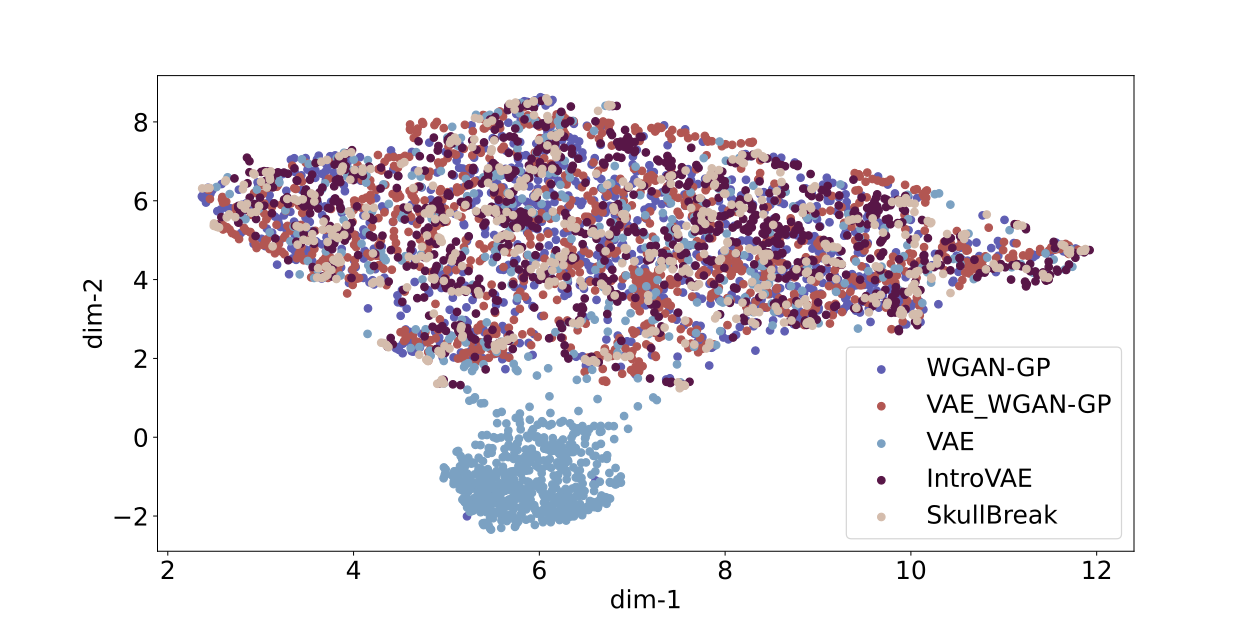}
		\subcaption{\label{umap}}
	\end{subfigure}
	\caption{\label{fig:exampleFromDataset} (\protect\subref{tsne}) t-SNE and (\protect\subref{umap}) UMAP projection of the real data and data generated by the various models.}
\end{figure}

\subsection{Qualitative analysis - latent manifold analysis}
Learned distribution can be also analyzed from a qualitative perspective. There is a strong intuition
behind this approach as it is possible to judge visually whether the synthetic skulls match the real ones in terms of shape, structure, defect orientation, etc. However, to perform this comparison in a more global manner that will jointly include all the real and generated data, it’s more appropriate to focus on the latent representations. It comes from a fact that if the data samples are similar in high dimensionality, they should be close to each other in low dimensions. This property can be viewed from both a global and local perspective, hence we analyze t-Distributed Stochastic Neighbor Embedding (t SNE)~\cite{Maaten2008}, together with Uniform Manifold Approximation and Projection (UMAP)~\cite{McInnes2018}. t-SNE pays more attention to the local structures' preservation and is presented in the Figure \ref{tsne}, while UMAP is mainly focused on maintaining the global structure \ref{umap}.

\begin{figure*}[h]
\begin{center}
    \includegraphics[width=\textwidth]{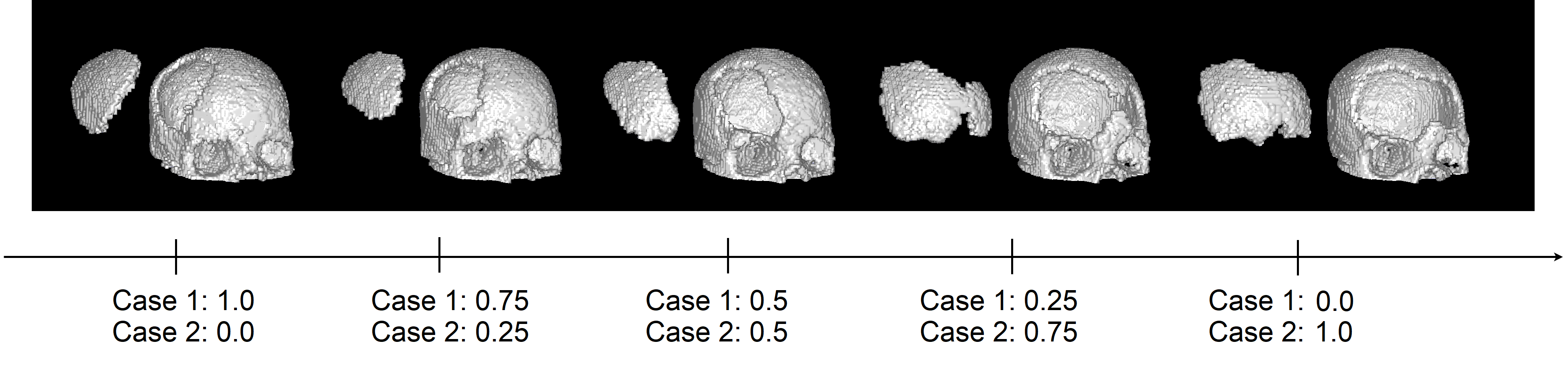}
    \caption{\label{fig:interpolations}Interpolation between two skulls in their latent space using VAE/WGAN-GP for a generation.}
\end{center}
\end{figure*}

\begin{figure*}[h]
\begin{center}
    \includegraphics[width=\textwidth]{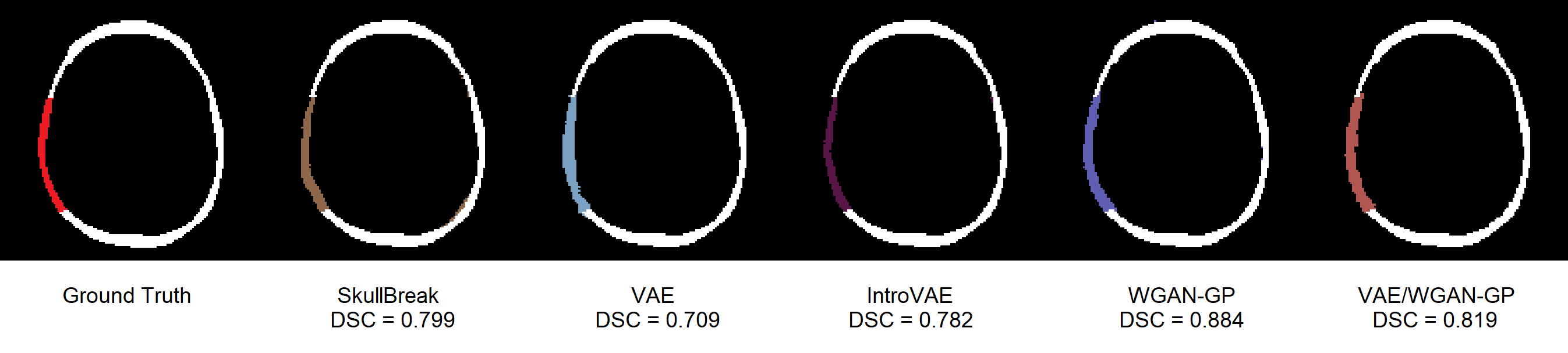}
    \caption{\label{fig:vizs}Results of defect segmentation on a skull from SkullBreak test set with the use of V-Nets trained on data from different sources.}
\end{center}
\end{figure*}

\section{Discussion}
This work studied and analyzed several crucial aspects of deep generative networks used for cranial implant design. The first and key takeaway is that with the use of the proposed methods it is possible to synthetically generate realistic defective skulls together with their compatible defects. The V-Net defect segmentation task produced the best results on SkullBreak test data when trained with the use of WGAN-GP and VAE/WGAN-GP generated skulls. The promising outcome of the quantitative analysis is that pretraining the WGAN-GP with the VAEs latent manifold for even just a few epochs, can have a very positive influence on the synthetic data quality and hence, improve defect segmentation results (higher Sørensen–Dice coefficient), and provide similar results as WGAN-GP, however with faster convergence. For each generative model, it is visible that the V-Net generalizes better into the real data as the amount of synthetic data increases. Thus, we find that the larger the size of the synthetic dataset is, the better the segmentation results on real data are, and less overfitting occurs. We also observe that for the relatively smaller synthetic datasets, the results of segmentation are worse in comparison to training on the real SkullBreak training set directly. It is caused by the fact that original train and test datasets are derived from the same data distribution, while synthetic samples are more heterogeneous, hence a larger amount is required to enable test set generalizability.  We have shown that training V-Nets with each of the proposed generative networks' large datasets (without using any real examples from the SkullBreak) results in better performance on the test data when compared to training only on the real skulls from the initial training set. The improvement in Sørensen–Dice coefficient is as large as 0.05 for WGAN-GP and 0.04 for VAE/WGAN-GP. Furthermore, we also observed that these networks achieved better results than simple models like the mentioned classic VAE, which didn't outperform training on real SkullBreak data and its latent representation stands apart from the rest. 
From the qualitative point of view, the latent manifold on which data from SkullBreak lies is accurately approximated by the proposed networks. It is clearly shown in the t-SNE and UMAP analysis, that both local and global structures preserve the SkullBreak data distribution and lie very close to it in the latent space. This is an informative property of these networks' generative processes, as it shows that they managed to learn the real data distribution and sample from it. Thus, they satisfy the main objective of training robust generative models. 

What's important we also note several limitations of our work, where one of them is the fact of training the generative networks on a lower resolution compared to the original resolution of the skulls. We believe that training on the higher resolution might positively influence the segmentation task and additionally be more useful for real medical studies, although it also requires more computational resources. We also note, that deeper dive into the hyperparameters of the networks, as well as training pipelines, might reduce the need for postprocessing steps and have a positive impact on the robustness of the models.

Bringing it all together, the work presents that with the use of deep generative networks, it is possible to generate highly realistic defective skulls which can be used to improve automatic cranial defect reconstruction. Moreover, the generated skulls fully capture the dynamics and heterogeneity of cranial defects, as the models correctly learn the distribution of real data and provide diverse latent spaces. In the future work, we will explore more recent generative models like diffusion~\cite{ho2020denoising} or latent diffusion frameworks~\cite{rombach2022high}, and increase the volumetric shape of the generated skulls to capture the fine details that may be important from the aesthetic point of view.

\section*{Acknowledgements}
The project was funded by The National Centre for Research and Development, Poland under Lider Grant no: LIDER13/0038/2022 (DeepImplant). We gratefully acknowledge Polish HPC infrastructure PLGrid support within computational grant no. PLG/2023/016239.

The preprint has not undergone peer review or any post-submission improvements or corrections. The Version of Record of this contribution will be published in (Information in future) and is available online at: (Information in future).

{\small
\bibliographystyle{ieee_fullname}
\bibliography{bibliography}
}

\end{document}